\documentclass[10pt,fleqn,twosides]{article}
\usepackage{mathrsfs}
\usepackage{latexsym}
\usepackage{tabmac}
\usepackage{amsthm}
\usepackage{amsopn,amsfonts,amssymb}
\usepackage{fancyhdr}
\usepackage{latexsym}
\fancypagestyle{plain}{ \fancyhead{} \fancyfoot{}

}

\newcommand{\cercle}[1]{\ensuremath{\setlength{\unitlength}{1ex}\begin{picture}(2.8,2.8)\put(1.4,0.7){\circle{2.8}\makebox(-5.6,0){#1}}\end{picture}}}
\newcommand{\tcercle}[1]{\ensuremath{\setlength{\unitlength}{1ex}\begin{picture}(2.8,2.8)\put(1.4,1.4){\circle{2.8}\makebox(-5.6,0){#1}}\end{picture}}}

\theoremstyle{definition}

\theoremstyle{remark}

\newcommand{\beq}{\begin{equation}}
\newcommand{\eeq}{\end{equation}}
\newcommand{\beqa}{\begin{eqnarray}}
\newcommand{\eeqa}{\end{eqnarray}}

\newcommand{\mc}{\mathcal}

\let\d\partial

\let\la\lambda
\let\La\Lambda

\let\ta\theta

\let\rw\rightarrow
\let\Rw\Rightarrow

\newcommand{\LL}{\ensuremath{\langle\!\langle}}
\newcommand{\RR}{\ensuremath{\rangle\!\rangle}}

\let\La\Lambda
\let\la\lambda
\let\ta\theta
\let\d\partial
\let\rw\rightarrow

\begin{document}

\title{Jack superpolynomials: physical and combinatorial definitions
     }

\author{Patrick Desrosiers\thanks{pdesrosi@phy.ulaval.ca} \\
\emph{D\'epartement de Physique, de G\'enie Physique et d'Optique},\\
Universit\'e Laval, \\
Qu\'ebec, Canada, G1K 7P4.
\and
Luc Lapointe\thanks{lapointe@inst-mat.utalca.cl }\\
\emph{Instituto de Matem\'atica y F\'{\i}sica},\cr
Universidad de Talca,\cr
Casilla 747, Talca, Chile.
\and
Pierre Mathieu\thanks{pmathieu@phy.ulaval.ca} \\
\emph{D\'epartement de Physique, de G\'enie Physique et d'Optique},\\
Universit\'e Laval, \\
Qu\'ebec, Canada, G1K 7P4.
}

\date{June 2004}

\maketitle

\begin{abstract}

Jack superpolynomials are  eigenfunctions of the supersymmetric extension
of the quantum trigonometric Calogero-Moser-Sutherland. They are
orthogonal with respect to the scalar
product, dubbed physical, that is naturally induced by this quantum-mechanical
problem. But Jack superpolynomials can also be defined more
combinatorially, starting
from the multiplicative bases of symmetric superpolynomials,  enforcing
orthogonality with respect to a one-parameter deformation of the
combinatorial scalar
product. Both constructions turns out to be equivalent. This provides strong
support for the correctness of the various underlying constructions
and for the pivotal role of Jack
superpolynomials in the theory of symmetric superpolynomials.
\footnote{To appear in the proceedings of the {\it XIII International
Colloquium on Integrable Systems and
Quantum Groups}, Czech. J . Phys., June 17-19 2004, Doppler
Institute, Czech Technical University, ed. by
C. Burdik}
\end{abstract}

\newpage

The aim of this contribution is to highlight some aspects of our work
\cite{DLM1,DLMa}, focussing on the equivalence between two totally
different approaches to the
construction of Jack superpolynomials: a physical one, in terms of an
eigenvalue problem in
supersymmetric quantum mechanics \cite{DLM1} and a more mathematical
definition linked to algebraic
combinatorics \cite{DLMa}, along  the lines of \cite{Stan}.  Most
references to original works are
omitted here and can be found in  these quoted articles.
Moreover,
the presentation is kept at a rather informal level.

\section{Jack superpolynomials as eigenfunctions of the stCMS model}

Jack superpolynomials are eigenfunctions of the supersymmetric
extension of the quantum trigonometric Calogero-Moser-Sutherland
(stCMS) Hamiltonian,  without its ground state
contribution.\footnote{We stress that Jack superpolynomials (also
called Jack polynomials in
superspace) do not appear to have any relation to the super-Jack
polynomials of \cite{VS},
  based on Lie superalgebras.} This model describes the
interaction of $N$ particles on a unit radius circle, with canonical variables
$x_i$ and
$ p_i= -\mathrm{i}\d /\d x_i$ (with $x_j$ subsequently replaced by $z_j=
\exp(\mathrm{i}x_j)$), together with their fermionic partners, the grassmannian
variables
$\ta_i$ (with
$\ta_i\ta_j=-\ta_j\ta_i$) and
$\ta_i^\dagger=\d/\d_{\ta_i}$:
    \begin{equation} \label{shjack}
\bar{\mc{H}}= \sum_i (z_i \partial_i)^2+\beta \sum_{i<j}\frac{ z_i+z_j}
{z_{ij}}(z_i \partial_i-z_j\partial_j)-2\beta\sum_{i<j}\frac{z_i
z_j}{z_{ij}^2}(1-\kappa_{ij}) \, .
\end{equation}
   The whole dependence
upon fermionic variables is contained in the factor $\kappa_{ij}$ which reads
\beq \label{deka}
\kappa_{ij}\equiv 1-\theta_{ij}\theta^\dagger_{ij}=
1-(\theta_{i}-\theta_j)(\partial_{\theta_i}-\partial_{\theta_j}).
\eeq
Remarkably, this is a fermionic-exchange operator,
i.e., $\kappa_{12}\ta_1\ta_2= \ta_2\ta_1$ ($=-\ta_1\ta_2)$.
That the supersymmetric extension is fully captured by the
introduction of a fermionic exchange
operator is
  a key
technical tool in the
integrability analysis.

The Jack superpolynomials are thus eigenfunctions of the operator
(\ref{shjack}). More precisely, the Jack superpolynomials
diagonalize the full set of $2N$ commuting conserved operators of
the stCMS model. This readily implies their orthogonality  with
respect to the `physical' scalar product : \beq
\label{physca}\langle
A(z,\theta),B(z,\theta)\rangle_{\beta}=\prod_{j}\frac{1}{2\pi
\mathrm{i}} \oint \frac{ dz_j}{z_j}\int
d\theta_j\,\theta_j\prod_{k\neq
l}\left(1-\frac{z_k}{z_l}\right)^\beta A(z,\theta)^*B(z,\theta)\,
,\eeq where the complex conjugation $*$ is defined
as\beq\label{defcomplex} z_j^*= 1/z_j\quad\mbox{and}\quad
(\theta_{i_1}\cdots\theta_{i_m})^*\theta_{i_1}\cdots\theta_{i_m}=1
\;.\eeq
   The integration over the
Grassmannian variable is the standard Berezin integration, i.e.,
$ \int d\theta=0$ and $\int d\theta \,\theta=1$.

\section{Jack superpolynomials as symmetric superfunctions}

Ordinary Jack polynomials are symmetric polynomials, i.e., invariant under the
action of the  operator
$K_{ij}$ that exchanges the variables $z_i$ and $z_j$:
\beq
            K_{ij}f(z_i, z_j)=f(z_j, z_i)K_{ij} \, .
\eeq
Jack superpolynomials are invariant under a generalization of this
condition, namely, under the simultaneous
exchange of the bosonic and the fermionic variables generated by:
\beq
{\cal K}_{ij}= K_{ij}\kappa_{ij} \, .
\eeq
where $\kappa_{ij}$ is defined in (\ref{deka}).
   Manifestly,
$\bar{\mc{H}}$ leaves invariant the space
of polynomials of a given degree in $z$ and a given degree in
$\theta$, being homogeneous in both sets of variables.
Eigenfunctions are thus
of the  form:
\begin{equation}
\mc{A}^{(m)} (z,\theta;\beta)=\sum_{1 \leq i_1< \ldots <i_m \leq
N}\theta_{i_1}\cdots\theta_{i_m} A^{(i_1\ldots i_m)}(z;\beta)\, ,
\quad m=0,1,\dots, N \, ,
\end{equation}
where $ A^{(i_1\ldots i_m)}$  is a homogeneous polynomial in  $ z$.
Note the simple dependence upon the
fermionic variables, which factorizes in  monomial prefactors, a manifest
consequence of their anticommuting nature.
Since the
eigenfunctions $\mc{A}^{(m)}$ are assumed to be invariant under the
action of the
     exchange operators
$\mc{K}_{ij}$ and given that the $\theta$ products are
antisymmetric, \emph{i.e.}, \beq \kappa_{jk} \,
\theta_{i_1}\cdots\theta_{i_m}=- \theta_{i_1}\cdots\theta_{i_m}
\quad\mbox{ if } \quad j,k \in \{i_1,\ldots, i_m\} \, , \eeq the
superpolynomials $ A^{i_1\ldots i_m}$ must  be partially
antisymmetric to ensure the complete symmetry of $\mc{A}^{(m)}$. In
fact, each polynomial $ A^{i_1\ldots i_m}$ is completely
antisymmetric in the variables $\{ z_{i_1},\ldots, z_{i_m}\}$, and
totally symmetric in the remaining variables $z\setminus
\{z_{i_1},\ldots, z_{i_m}\}$.

In the same way as symmetric
polynomials are indexed by
partitions, symmetric
superpolynomials are indexed by
{superpartitions}. A
superpartition is a sequence of non-negative integers
that generates two partitions separated by a semicolon: \beq
\Lambda=(\Lambda_1,\ldots,\Lambda_m;\Lambda_{m+1},\ldots,\Lambda_N)=
(\Lambda^a ; \Lambda^s), \eeq $\La^a$ being associated to an
antisymmetric function of the variables $\{ z_{i_1},\ldots, z_{i_m}\}$ (so that
$\Lambda_i>\Lambda_{i+1}$ for $  i=1, \ldots m-1$) and $\La^s$, to a symmetric
function of the variables $\{ z_{i_{m+1}},\ldots, z_{i_N}\}$, (i.e.,
$\Lambda_i \ge
\Lambda_{i+1}$ for $i\geq m+1$).
For $m=0$, the semicolon disappears and we recover the partition
$\Lambda^s$. The number $m$ is called the fermionic degree. The
bosonic degree of a superpartition is simply the sum of its parts.
For instance, the superpartitions of bosonic and fermionic degrees
3 and 1 respectively are: \beq (3;0), \quad (2;1),\quad (1;2), \quad
(1;1,1), \quad (0;3), \quad (0;2,1), \quad (0;1,1,1)\,. \eeq

    A superpartition $\La=(\La^a;\La^s)$ can also be
viewed as a partition (by reordering its parts) in which every part of
$\La^a$ is encircled. If a part ${\La^a}_j=b$ is equal to at
least one part of $\La^s$, then we circle
the leftmost $b$.  For instance,
\beq\Lambda=(3,0;4,3)
  \equiv(4,\cercle{3},3,\cercle{0})\,.\eeq
This allows
us to associate to each $\La$
a unique superdiagram, denoted by
$D[\La]$ in which the `fermionic rows' (circled parts) have an
additional circle at the end. We will also need to 
introduce the transposition
$\La'$ of $\La$,
which  is defined by the transposition of the corresponding
superdiagram, an operation
that manifestly preserves the fermionic degree (number of circles).
For instance, we have
\beq D([3,0;4,3] )=
{\tableau[scY]{&&&\\&&&\bl\tcercle{}\\&&\\\tcercle{}\bl\\ }}\, , \quad
D([3,0;4,3])^t=
{\tableau[scY]{&&&\bl\tcercle{}\\&&\\&&\\&\bl\tcercle{}\\ }}\, , \eeq
meaning that $(3,0;4,3)'= (3,1;3,3)$.

The Jack polynomials are most naturally defined in an expansion in
terms of monomial symmetric
functions. This is also true for their superextensions.
The monomial symmetric superpolynomials (or supermonomials)
are defined as:
\begin{equation}
m_{\Lambda}(z,\theta)={\sum_{\sigma\in S_{N}}}' \theta_{\sigma(1)}
\cdots \theta_{\sigma(m)}\, z_1^{\Lambda_{\sigma(1)}}  \cdots
z_m^{\Lambda_{\sigma(m)}} z_{m+1}^{\Lambda_{\sigma(m+1)}} \cdots
z_{N}^{\Lambda_{\sigma(N)}} \, , \end{equation} where $S_N$ is the
symmetric group, the prime indicates that the  summation is
restricted to distinct terms. For instance, for $N=3$ and
$\La=(3,1;2)$, we have
\beq m_{(3,1;2)}= \theta_1 \theta_2 (z_1^3z_2-z_1z_2^3)
z_3^2+\theta_1 \theta_3 (z_1^3z_3-z_1z_3^4)z_2^2 +\theta_2 \theta_3
(z_2^3z_3-z_2z_3^3)z_1^2\, . \eeq

We can now define the
Jack  superpolynomials as the
unique eigenfunctions of the supersymmetric Hamiltonian
     $\bar{\mc{H}}$,
\begin{equation}
\bar{\mc{H}}\, {J}_\Lambda^{(\beta)} (z,\theta)  =
\varepsilon_\Lambda {J}_{\Lambda}^{(\beta)}(z,\theta)\, ,
\end{equation}
(for some eigenvalue $\varepsilon_\Lambda$ that do not need to be specified)
     that can be decomposed triangularly in terms of the supermonomials:
\begin{equation}\label{trian}
{J}_\Lambda^{(\beta)} (z,\theta)=m_{\Lambda} (z,\theta)+\sum_{\Omega;\,
\Omega <  \Lambda}c_{\Lambda,\Omega}(\beta) m_{\Omega}(z,\theta)\, .
\end{equation}
and which are orthogonal with respect to the physical scalar product
(\ref{physca}):
\beq
   \langle
J^{(\beta)}_\La(z,\theta),J^{(\beta)}_\Omega(z,\theta)\rangle_{\beta}\propto
\delta_{\La,\Omega}\, . \eeq
   Observe that if we delete the
semi-colon of a superpartition, it becomes a composition (i.e., a
non-ordered sequences of
non-negative integers). These are naturally ordered by the Bruhat
ordering.
Here is a Jack superpolynomial of bosonic degree 2 and
fermionic degree 1:
\begin{eqnarray}
{J}^{(\beta)}_{(2;0)}= m_{(2;0)}+ \frac{\beta}{2+\beta}\, m_{(0;2)}+
\frac{2\beta}{2+\beta}\, m_{(1;1)}+
\frac{2\beta^2}{(1+\beta)(2+\beta)}\, m_{(0;1,1)}\, .
\end{eqnarray}
Obviously, in the limit $\beta\rightarrow 0$, the Jack
superpolynomial $J_\La^{(\beta)}(z,\ta)$ reduce to supermonomial $
m_\La(z,\ta)$.  Note also that for $m=0$, $J_\La^{(\beta)}(z,\ta)$ is
the ordinaty Jack polynomial $J_{\La^s}^{(\beta)}(z)$.

\section{Multiplicative  bases  of symmetric
superpolynomials}

The Jack superpolynomials $J_\La^{(\beta)}$'s and the supermonomials $m_\La$'s
provide two bases for the space of symmetric superpolynomials. There
are other natural candidate  bases, namely those that would result
from the superextension of the elementary $e_n$, homogeneous $h_n$
and power sum $p_n$ symmetric functions, naturally defined by their
generating functions: \beq \sum_{n\geq 0} e_n\, t^n=\prod_{i\geq
1}(1+z_it)\; ,\quad \sum_{n\geq 0} h_n\, t^n=\prod_{i\geq
1}\frac{1}{(1-z_it)}\;,\quad \sum_{n\geq 1} p_n\, t^n=\prod_{i\geq
1}\frac{z_i t}{(1-z_it) }\,. \eeq The natural way of extending these
symmetric functions is to make the following replacement in their
generating functions: \beq
   tz_i\rw tz_i+\tau
\ta_i \, ,\eeq where $\tau$ is a constant anticommuting parameter
($\tau^2=0)$. That  makes the resulting functions manifestly
invariant under the action of ${\cal K}_{ij}$. Denote by
$\tilde{e}_n$, $\tilde{h}_n$ and $\tilde{p}_n$ the coefficient of
$\tau t^n$ in each modified generating function. For instance, we
have \beq\prod_{i\geq 1}(1+z_it+\tau \ta_i)= \sum_{n\geq 0}
t^n[e_n+\tau \tilde{e}_n] \eeq with \beq e_n=\sum_{1\leq
i_1<\ldots<i_n\leq N} z_{i_1}\cdots z_{i_n},\qquad
\tilde{e}_{n-1}=\sum_{1\leq j\neq i_1,\ldots,i_{n-1}\leq N\atop
1\leq i_1<\ldots<i_{n-1}\leq N}\theta_j z_{i_1}\cdots
z_{i_{n-1}}\eeq with $e_0=1$ and $
e_{n}=\tilde{e}_{n-1}=0\quad\forall n>N$. Similarly, we find \beq
p_n=\sum_{i=1}^N z_i^n,\quad \tilde{p}_{n-1}=\sum_{i=1}^N \theta_i
z_i^{n-1}\, . \eeq The sets $\{g_n,\tilde{g_n}\}$ for
$g_n\in\{e_n,h_n,p_n\}$ form  new multiplicative bases for symmetric
superpolynomials, e.g.: \beq p_\Lambda=\tilde{p}_{\Lambda_1}\cdots
\tilde{p}_{\Lambda_m}p_{\Lambda_{m+1}}\cdots p_{\Lambda_N}\, .\eeq

Recall also the Cauchy formula, from which a `combinatorial' scalar
$\LL \;|\;\RR $ product can be defined: \beq
\prod_{i,j}\frac{1}{(1-z_iw_j)}= \sum_\la a_\la^{-1} p_\la(z)\,
p_\la(w) \quad \Rw \quad \LL p_\la\,| \,  p_\mu \RR=
a_\la\,\delta_{\la,\mu}\, ,\eeq where for $\la=
(1^{m_1}2^{m_2}\cdots)$, $a_\la= \prod_i i^{m_i} m_i!$. Again, this
can be extended naturally as follows: \beq \label{supca} \prod_{
i,j}\frac{1}{1-z_iw_j-\theta_i\phi_j}= \sum_{\La} a_\La^{-1}\,
p_\La(z,\theta)\, p_\La(w,\phi)\quad \Rw \quad  \LL p_\La\,| \,
p_\Omega \RR= a_\La \delta_{\La,\Omega}\, ,\eeq where $ a_\La=
(-1)^{m(m-1)/2}\, a_{\La^s} $, $m$ being the fermionic degree of
$\La$.
This product can be deformed  by considereing the $\beta$-th power
of the product in (\ref{supca}): \beq \LL p_\La\,| \, p_\Omega
\RR_\beta= \beta^{-\ell(\La)}a_\La \delta_{\La,\Omega}\, ,\eeq
where $\ell(\La)$ is the length of the superpartitions: $\ell(\La)=
m+\ell(\La^s)$. It is known that the Jack polynomials are orthogonal
with respect to both the physical scalar product (\ref{physca}) and
the above $\beta$-deformed combinatorial product, evaluated at
$m=0$. This turns out to be also true for the Jack superpolynomials.
To be precise, if we expand $J_\La^{(\beta)}$ in the basis of the super
power-sums $p_\La$'s,
then we check that \beq \LL  {J_\La^{(\beta)}}\,| \, {J_\Omega^{(\beta)}}
\RR_\beta\propto \delta_{\La,\Omega}\, .\eeq Actually, the Jack
superpolynomials can be uniquely reconstructed in this way, together
with the usual triangularity requirement.

Another contact between $J_\La^{(\beta)}$ and the combinatorial bases is the
following relation that also generalizes a well-known limiting case
of the Jack polynomials \beq \label{jvse} (-1)^{m(m-1)/2}
\lim_{\beta\rightarrow\infty} \, J_\La^{(\beta)}(z,\ta) =
e_{\La'}(z,\ta)\, . \eeq  Here is a simple illustration of
(\ref{jvse}) for $N=3$: \beq J_{(2,0;0)}^{(\beta)}=m_{(2,0;0)}+
\frac{\beta}{1+\beta} \, m_{(1,0;1)}\quad \Rw\quad
\lim_{\beta\rw\infty}J_{(2,0;0)}^{(\beta)}= m_{(2,0;0)}+m_{(1,0;1)}\, . \eeq
We have $(2,0;0)'=(1,0;1)$ and \beq e_{(1,0;1)}=
[\ta_1(z_2+z_3)+\ta_2(z_1+z_3)+\ta_3(z_1+z_2)]\,
[\ta_1+\ta_2+\ta_3]\, [z_1z_2+z_1z_3+z_2z_3]\, ,\eeq where
$e_{(1,0;1)}= {\tilde e}_1{\tilde e}_0 {e}_1$.  It is simple to see
that the last two expressions are equivalent: for this it suffices
to check the coefficient of $\ta_1\ta_2$ which is
$-(z_1-z_2)(z_1+z_2+z_3)$.

\section{Conclusion}

  We should point out that the first
approach (the  physical) presented has already been extended to the
construction of the
Hermite generalized superpolynomials, eigenfunctions of the rational
supersymmetric CMS model with confinement \cite{DLMb}. However, the
supersymmetric version of the Ruijsenaars model is still missing.
The Macdonald superpolynomials, their would-be eigenfunctions, can
nevertheless be constructed combinatorially, along the above lines.

      \vskip0.3cm
\noindent {\bf ACKNOWLEDGMENTS}

This work was  supported by NSERC and FONDECYT
grant \#1030114. P.D. is
grateful to the Fondation J.A.-Vincent for a student fellowship.


\begin{thebibliography}{99}
\addcontentsline{toc}{section}{References}

           \bibitem{DLM1}
P.~Desrosiers, L.~Lapointe and P.~Mathieu, Nucl. Phys. {\bf B606}
(2001) 547, hep-th/0103178; Commun. Math. Phys. {\bf 233} (2003)
383, hep-th/0105107; Commun. Math. Phys. {\bf 242} (2003) 331,
hep-th/0105107. See also the review:
hep-th/0210190.


\bibitem{DLMa}
P.~Desrosiers, L.~Lapointe and P.~Mathieu, \emph{Symmetric functions in
superspace}, in preparation.

\bibitem{Stan}
               R. P. Stanley, Adv. Math. {\bf 77}  (1988) 76.




\bibitem{VS}
A.N.~Sergeev and A.P. Veselov, Commun. Math. Phys. {\bf 245} (2004)
249-278.


\bibitem{DLMb}
P.~Desrosiers, L.~Lapointe and P.~Mathieu, Nucl.
Phys. {\bf B674} (2003) 615; J. Phys. A: Math. Gen. {\bf 37} (2004) 1251.






\end{thebibliography}
\end{document}